# Sound-first immersive training for blind and low-vision learners: A simulation flow for safe, standardized orientation, mobility, and daily living practice

Daniel A. Muñoz[1]

[1]Department of Interactive Media, School of communication, Hong Kong Baptist University, 999077, Hong Kong SAR

**\*Correspondence to:** Daniel A. Muñoz, Department of Interactive Media, School of communication, Hong Kong Baptist University, Hong Kong SAR, 999077, damunoz@hkbu.edu.hk

**Abstract**

Orientation and mobility (O&M) instruction for blind and low-vision learners is effective but difficult to standardize and repeat at scale due to the reliance on instructor availability, physical mock-ups, and variable real-world outdoor conditions. This Technical Note presents a sound-first immersive training flow that uses spatial audio and sonification as the primary channel for action and feedback in pre-street O&M and daily-living practice. The approach specifies parameterized scenario templates (e.g., signalized street crossing, public transport boarding, and kitchen tasks), a compact and consistent cue vocabulary with clear spectral placement and timing to mitigate masking, and a lightweight safety protocol enabling graded exposure, content warnings, seated starts, opt-outs, and structured debriefs. The system assumes a head-mounted device with high-quality binaural rendering and head tracking; 3D scene geometry is used as an invisible scaffold to anchor sources, trigger events, define risk/guidance volumes, and govern physically plausible motion without visuals. Session difficulty is shaped via cue density, event tempo, and task complexity while preserving cue consistency to promote transfer across scenarios. The specification aims to enable safe repetition, reduce instructor burden, and support clearer standards across rehabilitation centers, aligning with evidence that audio-first interaction is essential for blind and visually impaired users and addressing gaps in HRTF personalization, evaluation standards, and accessibility integration. Although no behavioral outcomes are reported here, this implementable flow consolidates auditory science with center-ready design, offering a pragmatic foundation for standardized evaluation and future comparative studies.

## 1. Introduction

Orientation and mobility (O&M) programs give independence to VI/B (blind and visually impaired) students through verbal guidance, tactile or physical models, and graded exposure to real environments. This instructor driven model is effective but constrained by the facility space, for instance, staging indoor bus stops or mock up kitchens to develop familiarity and train.

This approach is also constrained by weather and timing, by crowding and traffic variability, and by safety concerns during early stages of training. In cooking instruction, learners must calibrate to heat, the characteristic sound of hot oil, and smell; these are skills that require repeated practice and careful supervision but are difficult to stage safely and consistently. As a result, many learners face few opportunities for standardized, independent practice at a level of difficulty that is appropriate to their current abilities.

Immersive technologies can complement instructor-driven training when designed audio-first. Calibrated spatial sound can support orientation, cue timing, and corrective action under controlled, repeatable conditions. However, many immersive environments remain sight-first, where audio is layered late as prompts over visual workflows as result of an after-inclusion practice. When audio is not the primary channel, cues often become inconsistent across scenarios, learners are nudged to follow instructions rather than develop robust auditory strategies, and cognitive effort increases to interpret mismatched mappings. A recent review concludes that audio-first design is not optional for blind and visually impaired users and identifies gaps in HRTF (head related transfer functions) personalization, evaluation standards, and integration of accessibility in immersive design pipelines [1]. Foundational spatial audio research demonstrates how rendering choices and room acoustics shape localization, externalization, and orientation [2–4], while nonvisual interface studies show that stable, semantically consistent mappings improve learnability and transfer [5,6]. This evidence motivates a structured sound-first training flow that is safe, standardized, and repeatable before learners return to the street, the bus stop, or the kitchen.

## 1.1 Background and problem

The present training ecosystem confronts a persistent tension between realism and repeatability. Real streets, stations, and kitchens change from session to session: ambient noise fluctuates, event timing is irregular, and rare but critical scenarios, for example late amber crossings, misaligned bus doors, or sudden oil popping, are unpredictable (see analysis on Figure 1 and 2). Although this variability is valuable for ecological validity, it undermines controlled progression and the assignment of independent practice calibrated to learner readiness [7,8]. Furthermore, when immersive tools are designed with visual primacy, audio is reduced to guidance overlays. Learners can complete tasks by following prompts, but they do not necessarily acquire stable auditory strategies that generalize; inconsistent mappings between sound and action inflate workload and increase errors [5,6,9]. Safety and workload management add a third challenge. Without explicit control of loudness, cue density, and masking, which is the phenomenon by which overlapping sounds in time and frequency obscure one another, auditory scenes become tiring and hard to parse, shortening sessions and diminishing learning gains. Spatial rendering and room effects further affect localization and externalization, which are crucial for nonvisual orientation [2,3]. Evidence from perceptual learning and rehabilitation also indicates that coupling auditory cues to action strengthens spatial representations and supports mobility skill acquisition [10,11]. Ethical guidance for immersive experiences emphasizes graded exposure, comfort, and informed consent, which support the necessity to manage intensity and complexity over time [12]. Addressing these constraints calls for a standardized, sound-first training flow that scaffolds auditory discrimination, orientation, and decision timing under well-controlled conditions.

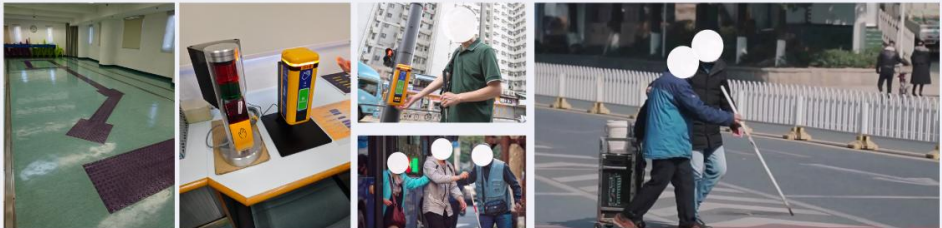

**Street and outdoor training scenarios**

**Current Training** — From Indoor practice directly to the street

1. Indoor crossing training using floor bumpers
2. Hong Kong Traffic light Auditive devices fo visually impaired users
3, 4 and 5. Visually Impaired users training in the street

1. Indoor crossing practice uses floor bumpers and fixed layouts; learners transition directly from this indoor practice to the street with limited control over noise, timing, and traffic density
2. Instruction relies on live environments and standalone devices such as Hong Kong traffic-light auditory units for visually impaired users; cues are heard in isolation and vary by location.
3. Field sessions with visually impaired users on the street provide real experience but expose learners to potential risks and allow few safe repetitions.

**Kitchen Training**

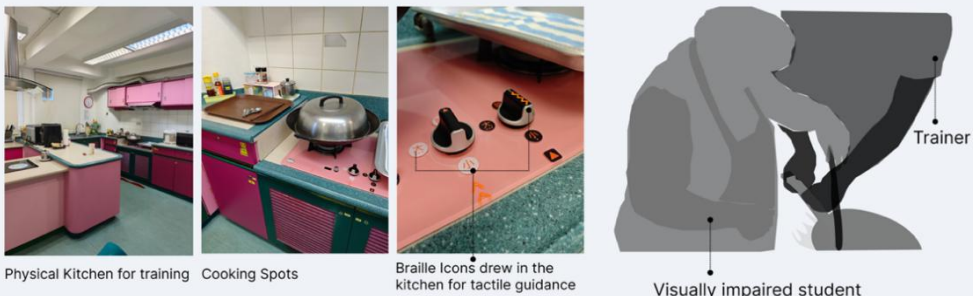

**Current Training**

Physical Kitchen for training • Cooking Spots • Braille Icons drew in the kitchen for tactile guidance • Trainer • Visually impaired student

Photos Taken by PI at the Hong Kong Society for the Blind (with permission). The Diagrams are PI elaboration

1. Instructor guides tactile and olfactory exploration; most task information is delivered through verbal description (Physical kitchen for training; Braille icons drawn in the kitchen for tactile guidance).
2. Heat and cutting tasks require close supervision, limiting early practice and constraining scenario variety (Cooking spots).
3. When frying, students must learn the smell of hot oil and the characteristic sounds, but practice opportunities to train acoustic and olfactory senses are limited and hard to stage safely.
4. Few opportunities for repetition, leaving limited time for adaptation and habit formation.

**Figure 1.** In the top, Analysis of Current training in outdoor scenarios for visually impaired and blind users

**Figure 2.** In the bottom Analysis of Current training in kitchen scenario for visually impaired and blind users

### 1.2 Empathic Computing framing and Operationalization

This work is situated within empathic computing, which we take as the design and evaluation of interactive systems that explicitly sense, model, and respond to users' perceptual, cognitive, and affective states to reduce harm, enable agency, and improve experience quality. In the context of orientation and mobility (O&M) for blind and low-vision learners, empathic design requires moving beyond generic accessibility toward responses tailored to auditory load, uncertainty, anxiety, and risk perception during graded exposure. Concretely, the proposed sound-first flow operationalizes empathy through four mechanisms:

1. State-aware workload shaping: Session parameters (cue density, event tempo, task complexity) are tuned to anticipate auditory masking and cognitive load, lowering error costs early and increasing complexity only as skills stabilize. This meets learners "where they are," rather than enforcing fixed difficulty.

2. Safety-by-design as affective scaffolding: Content warnings, seated starts, opt-out affordances, and mid-session check-ins explicitly address anticipatory anxiety and fatigue in early sessions. These are framed as standing rights, not exceptions, to preserve autonomy.

3. Semantic consistency for trust and transfer: A compact, stable cue vocabulary and congruent mappings reduce interpretive effort, mitigate surprise, and build confidence. Predictability is treated as an empathic property because it lowers vigilance burden.

4. Respectful sensing without intrusion: We use head tracking and scene geometry to stabilize spatialization and provide guidance without requiring visual attention or continuous corrective prompts. Guidance is minimally salient and directional, supporting dignity and independence rather than supervision.

These principles integrate psychoacoustics and nonvisual HCI with empathic intent: not only to make tasks possible, but to make practice psychologically safer, more legible, and more sustainable over time.

### 2. Methods

### 2.1 Overview of the Proposed Approach: a sound-first learning flow

We propose a practical method comprising scenario templates with adjustable parameters, a compact set of consistent interaction patterns and cue-design rules, a lightweight safety protocol, and the audio-only use of 3D scene information for triggering events and student guidance. The method assumes a head-mounted device with high-quality binaural rendering and head tracking to stabilize localization across head movements [4]. Rather than rendering visuals, the system uses 3D geometry and motion paths as an invisible scaffold, here we anchor spatial sound sources and occlusion (for example, a bus partially occluded by a shelter), triggers events as the learner moves through space (such as entering a curb-aligned zone), defines risk and guidance volumes (heat zones around burners or safe alignment zones at crossings), and drives moving sound objects with physically plausible kinematics (for example, a bicycle passing at a controlled lateral offset and speed). This approach maintains precise spatial cues while keeping the learner squarely in the auditory modality.

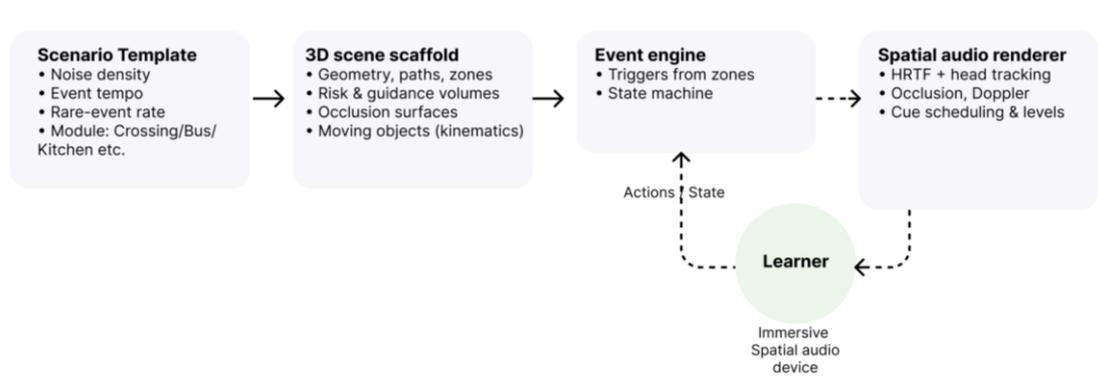

**Figure 3.** Proposed flow of sound first learning

### 2.1.1 Empathic design layer

To embed empathic responsiveness without adding intrusive sensing, we treat difficulty parameters as proxies for user state. Early-session defaults minimize concurrent sources and temporal crowding; recovery cues are brief, directional, and non-moralizing; confirmations emphasize state progress ("aligned," "safe gap confirmed") rather than failure. Session scripts incorporate scheduled micro-pauses for self-assessment and instructor check-ins, with an always-available opt-out and rapid de-escalation to quieter scenes. These choices are not ancillary "comfort features" but part of an empathic contract: the system proactively limits overload, acknowledges vulnerability during pre-street stages, and preserves learner autonomy while cultivating robust auditory strategies that generalize.

### 2.2 Cue Design and Masking Management

To support perception under realistic conditions, cue design explicitly manages masking and spectral placement. Masking occurs when sounds overlap in time and frequency, making important signals harder to hear, which is often a critical point in VI/B distinguishability training; here initially we place cues in frequency bands with less competition and schedule them to avoid overlap with dominant environmental sounds such as engine noise or fry sizzling, thereby improving detectability without increasing volume [13]. At this point short, semantically congruent confirmations such as brief tones or earcons (sound icons) that unambiguously mark state changes such as "aligned with crosswalk centreline" or "burner off" would reduce uncertainty without breaking task flow [5]. During this training, the error correction is guided by minimal-salience spatial signals conveying direction and magnitude of deviation, facilitating rapid recovery without compromising user autonomy. The workload is managed by shaping session difficulty through three factors: cue density, event tempo, and task complexity. Early sessions present fewer concurrent sources with slower timing; later sessions introduce denser noise fields, tighter crossing gaps, and rare events once core skills stabilize. Throughout, the cue vocabulary remains constant so that learned mappings transfer across scenarios. In this form the training increases cognitive complexity slowly giving space to students to practice distinguishability in a safe simulated environment.

**Sound First Immersive Approach**

Immersive sound simulation in a safe indoor space — Spatial sound and three dimensional information

Skill-building loop: Trial → Feedback → Adjust → Retry (Repeatable, Consistent feedback, Adaptive pacing)

Physical space with haptic feedback

Immersive Spatial audio device capable of augmenting digital information in physical world | Audio Prompts and spatial Audio guide | Proximity Audio Cue | Real sound from sources using spatial sound

Photo Taken by PI at the Hong Kong Society for the Blind (with permission). The Diagrams are PI elaboration

1. An immersive sound simulation in a safe indoor space reproduces real sound sources with spatial, three-dimensional audio, allowing repeated practice without exposure to traffic hazards
2. Spatial audio guidance and concise prompts augment the physical world via an immersive spatial-audio device, delivering consistent phase cues, gap-judgment training, and route-planning feedback.
3. Difficulty progresses systematically: increase masking noise, vary signal timing, add or remove hazards, and rehearse rare events, while preserving familiarity and safe room for error and correction.

**Figure 4.** Proposed immersive audio simulation and training system for learning outdoor situations.

The street-crossing module illustrates these principles in Figure 4. It parameterizes ambient noise density, vehicle approach profiles, signal timing variability, and occasional rare events. Learner's practice orienting their bodies, aligning with tactile indicators and curb ramps, judging safe gaps, and committing to cross at the appropriate moment. Spatialized auditory landmarks and concise state confirmations provide consistent, repeatable feedback. The public transport module similarly controls bus approach direction and speed, door alignment, and queue density, training the identification of the correct vehicle, alignment for boarding, and turn-taking; here, motion-consistent sonification (including appropriate Doppler cues) supports timing and trajectory judgments.

The kitchen module (see Figure 5) adapts these ideas to daily living. Heat proximity cues and risk-zone earcons define safe reach envelopes around burners; utensil sonification and oil temperature sound states provide scalable practice in sensing and timing; and rare events, such as an oil pop, can be introduced safely to build recognition. This module enables earlier, repeatable practice on auditory and procedural elements without burn or cut risk and later complements olfactory calibration and tactile exploration in the physical kitchen.

Safety and comfort are handled through a lightweight protocol: brief pre-screening and clear session briefings, content warnings for intense soundscapes, seated starts during initial sessions, immediate opt-out availability, short mid-session check-ins, and structured debriefs. These measures support comfort and graded exposure while minimizing additional burdens on instructors [12]. Because scenarios are parameterized and the cue vocabulary is fixed, the method can be standardized across centres and tuned to local soundscapes without losing comparability.

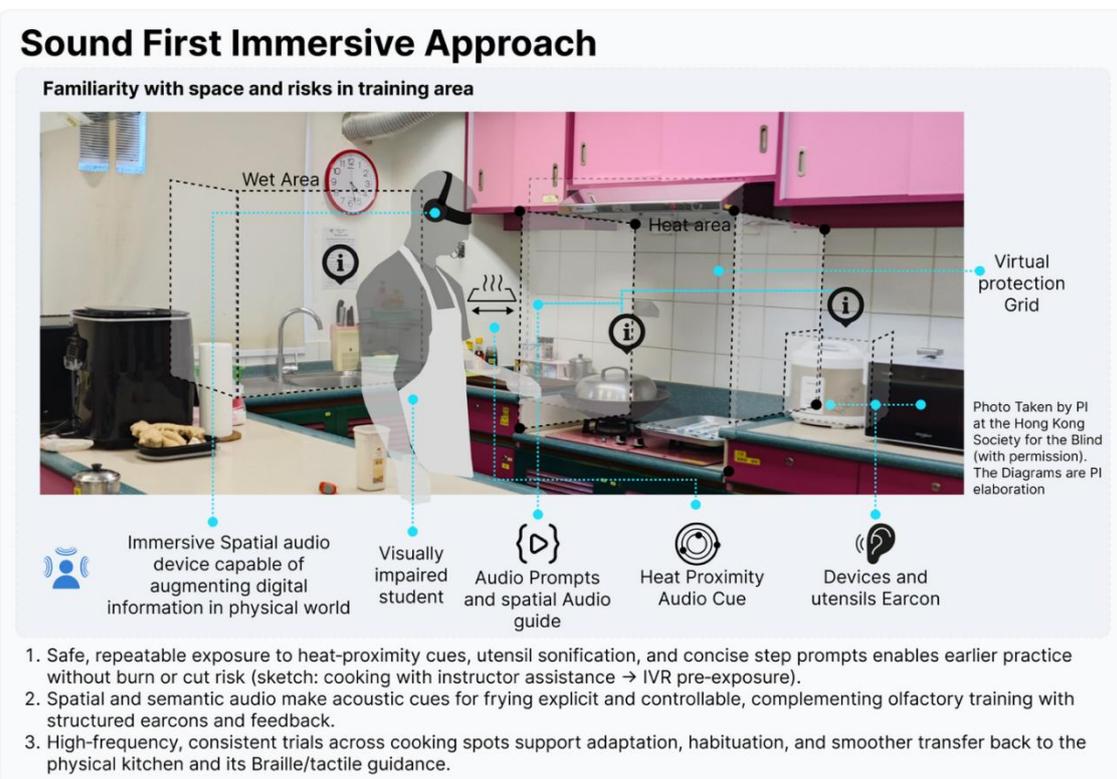

**Figure 5.** Proposed sound first immersive approach for familiarization and safe learning in cooking and indoor environments.

### 3. Discussion

The rationale for a sound-first flow is threefold. First, detectability and orientation improve when spatial cues are calibrated and masking is managed; a large body of psychoacoustic and spatial hearing research explains how interaural time and level differences, spectral shaping, and room interactions affect localization and externalization [2,3]. Second, learnability increases with consistent mappings between sounds and actions; nonvisual interface research shows that stable, semantically meaningful cues support faster acquisition and better transfer than ad hoc prompts[5,6]. Third, repetition under controlled variation helps learners calibrate to environmental sounds such as the traffic phases or vehicle approaches, before exposure to real scenarios, supporting safer transfer[10,11]. This method also directly answers gaps identified in the immersive sound literature for blind and visually impaired users, particularly the need to embed accessibility from the outset and to move toward clearer evaluation standards [1]. While this short communication does not report behavioural data, the flow is designed for pragmatic evaluation within centres, focusing on progress over time under standardized scenarios and leaving between-system comparisons to future work.

Empathic computing emphasizes system behaviors that read, respect, and respond to user states while safeguarding agency. In immersive accessibility, this implies more than providing alternative modalities: it requires calibrated control of workload, predictability, and risk signaling to reduce anxiety and fatigue and to support self-efficacy. The present contribution advances this agenda by specifying a sound-first training flow in which empathy is operationalized through anticipatory workload management, safety-by-design protocols, and a stable cue lexicon that reduces interpretive friction. Rather than inferring affect via biosignals, we adopt a pragmatic stance that couples psychoacoustic constraints and training progression to likely user states in early O&M practice. This makes the approach deployable in centers today while remaining compatible with future sensing (e.g., optional heart rate, speech prosody, or pause-frequency monitors) should evaluation warrant finer-grained adaptation. In this framing, empathy is not an afterthought layered over accessibility; it is a first-order design requirement that shapes scenario parameters, error recovery, and confirmations to protect dignity, foster trust, and support durable skill transfer

## 4. Conclusion

A sound-first, audio-only immersive training flow can make early O&M and daily living practice safer, more standardized, and more repeatable for blind and low-vision learners. By foregrounding spatial audio, managing masking and workload, and using 3D scene information to drive interaction without visuals, the approach consolidates established auditory science with practical, scenario design. It complements instructor led teaching, builds stable auditory strategies, and offers a clear path toward evaluable, accessible immersive training.

By explicitly framing difficulty control, safety protocols, and cue consistency as empathic responses to learner perceptual and emotional states, the system aligns with the core aims of empathic computing: reduce harm, respect autonomy, and enhance the quality of experience. The result is not only an accessible simulation but an empathically structured training flow that anticipates overload, normalizes graded exposure, and communicates state in clear, non-stigmatizing ways. This orientation elevates the contribution from modality substitution to empathic skill scaffolding, strengthening its fit with the journal's scope.

**Declarations**

**Acknowledgments**
Authors thank to the Hong Kong Society for the Blind (HKSB): instructor interviews, field and training visits, as well as authorization for photos and videos in the rehabilitation centre.

**Authors contribution**
The author contributed solely to the article.

**Conflicts of interest**
The authors declare no conflicts of interest.

**Ethical approval**
Not applicable.

**Consent to participate**
Not applicable.

**Consent for publication**
Not applicable.

**Availability of data and materials**
Not applicable.

# References

1. Muñoz DA. Inclusion Through Sound: A Systematic Review of Spatial Audio, Sonification, and Interaction Design in Immersive Technologies for Blind and Visually Impaired Users. In: *Human Factors in Design, Engineering, and Computing*. Vol 199. AHFE Open Acces; 2025. doi:10.54941/ahfe1006916

2. Begault DR. *3-D Sound for Virtual Reality and Multimedia*. Nachdr. AP Professional; 1997.

3. Blauert J, Blauert J. *Spatial Hearing: The Psychophysics of Human Sound Localization*. Rev. ed., 6. printing. MIT Press; 2008.

4. Wenzel EM, Arruda M, Kistler DJ, Wightman FL. Localization using nonindividualized head-related transfer functions. *The Journal of the Acoustical Society of America*. 1993;94(1):111-123. doi:10.1121/1.407089

5. Brewster S. Providing a structured method for integrating non-speech audio into human-computer interfaces.

6. Walker BN, Kramer G. Mappings and metaphors in auditory displays: An experimental assessment. *ACM Trans Appl Percept*. 2005;2(4):407-412. doi:10.1145/1101530.1101534

7. Foreman N. Spatial Cognition and Its Facilitation in Special Populations. In: *Applied Spatial Cognition: From Research to Cognitive Technology*. Lawrence Erlbaum Associates Publishers; 2007:129-177.

8. Ishikawa T. *Human Spatial Cognition and Experience: Mind in the World, World in the Mind*. Routledge; 2020. doi:10.4324/9781351251297

9. Brewster SA. Using nonspeech sounds to provide navigation cues. *ACM Transactions Computer-Human Interactions*. 1998;5(3):224-259. doi:10.1145/292834.292839

10. Rosenblum LD. *See What I'm Saying: The Extraordinary Powers of Our Five Senses*. 1st ed. W. W. Norton & Company, Incorporated; 2011.

11. Thaler L, Arnott SR, Goodale MA. Neural Correlates of Natural Human Echolocation in Early and Late Blind Echolocation Experts. Burr DC, ed. *PLoS ONE*. 2011;6(5):e20162. doi:10.1371/journal.pone.0020162

12. Cutchin S, Southgate E, Fails JA, Oliveira Da Silva MM. Workshop: IEEE VR KELVAR Workshop: K-12+ Embodied Learning through Virtual and Augmented Reality (8th Annual Workshop). In: *2023 IEEE Conference on Virtual Reality and 3D User Interfaces Abstracts and Workshops (VRW)*. IEEE; 2023:507-507. doi:10.1109/VRW58643.2023.00110

13. Moore BCJ. *An Introduction to the Psychology of Hearing*. 6. Aufl. Brill; 2013.